\documentclass{article}%
\usepackage{amssymb}%
\usepackage{amsmath}%
\usepackage{amsfonts}%
\usepackage{graphicx}

\begin{document}

\author{Vladimir K. Petrov\thanks{ E-mail address: vkpetrov@yandex.ru}}
\title{\textbf{Asymptotic series for distributions}}
\date{\textit{N. N. Bogolyubov Institute for Theoretical Physics}\\
\textit{\ National Academy of Science of Ukraine}\\
\textit{\ 252143 Kiev, Ukraine. }\textrm{02.02.2004}}
\maketitle

\begin{abstract}
Asymptotic expansions for a wide class of distribution are studied. Simple
method for the computation of the series coefficients is suggested. The case when
regularization parameter of distribution depends on the asymptotic parameter is considered.

\end{abstract}

\section{Introduction}

A method of asymptotic series expansion\ of $e^{\pm i\tau x}f\left(  x\right)
$ for $\tau\rightarrow\infty$ was suggested in \cite{br-shir,brych}. An
exhaustive investigation of the power type distributions, i.e. distributions
of a form $f\left(  x\right)  \sim\left(  x\pm i0\right)  ^{\lambda}\ln
^{m}\left(  x\pm i0\right)  $\ has been undertaken and the results were
presented in \cite{brych,brych-prud}. In this paper we extend this method to
the family of distributions specified below. Furthermore, we consider the case
when the regularization parameter in $f\left(  x\right)  $\ depends on $\tau
$\ and present resulting changes in the series expansion.

Recall that tempered distribution is defined as functional $\left\langle
F\left(  t\right)  ,\phi\left(  t\right)  \right\rangle $ on fast decreasing
test functions $\phi\left(  t\right)  \in S$ \cite{gel-shil,bremermann}.
Distribution $F\left(  t\right)  $ is tempered ($F\left(  t\right)  \in
S^{\prime}$), if and only if it is a finite order derivative of some
continuous tempered function $G\left(  t\right)  $, i.e. there exists some
finite $n$ and $\sigma$ such that
\begin{equation}
\left\vert G\left(  t\right)  \right\vert <\left\vert t\right\vert ^{\sigma
};\qquad\left\vert t\right\vert \rightarrow\infty,\label{cond_t}%
\end{equation}
and
\begin{equation}
F\left(  t\right)  =G^{\left(  n\right)  }\left(  t\right)  .\label{cond_n}%
\end{equation}
Fourier transform of any tempered distribution is a tempered distribution too.

Further it will be necessary to interchange the order of the integration in
the Fourier transform and taking the limit. To ensure uniform convergence of
the Fourier integral%
\begin{equation}
f\left(  x\right)  \triangleq%
%TCIMACRO{\tciFourier}%
%BeginExpansion
\mathcal{F}%
%EndExpansion
\left[  F,x\right]  \equiv\int_{-\infty}^{\infty}F\left(  t\right)
\exp\left\{  ixt\right\}  dt;\quad F\left(  t\right)  \in S^{\prime},\label{F}%
\end{equation}
necessary for such interchange, we apply the Abel-Poisson
regularization,\ i.e. $\left(  \ref{F}\right)  $ will be interpreted as
\begin{equation}
f\left(  x\right)  =\int_{0}^{\infty}F\left(  t\right)  \exp\left\{  i\left(
x+i\varepsilon\right)  t\right\}  dt+\int_{-\infty}^{0}F\left(  t\right)
\exp\left\{  i\left(  x-i\varepsilon\right)  t\right\}  dt,\label{AP}%
\end{equation}
where $\varepsilon$ is infinitesimal positive parameter which finally tends to
zero unless specified otherwise. 

It is known (see e.g. \cite{vladimirov}) that any tempered distribution
$F\left(  t\right)  $ allows an analytical representation%
\begin{equation}
f\left(  x\right)  =f_{+}\left(  x+i\varepsilon\right)  -f_{-}\left(
x-i\varepsilon\right)  ,\label{inv-AP}%
\end{equation}
where $f_{\pm}\left(  x\right)  $ are analytical functions in the upper/lower
complex half-plane $x$. In particular, $f_{\pm}\left(  x\right)  $ may be
chosen as%

\begin{equation}
f_{+}\left(  x\right)  =\int_{0}^{\infty}F\left(  t\right)  \exp\left\{
itx\right\}  dt\label{inv F+}%
\end{equation}
and%
\begin{equation}
f_{-}\left(  x\right)  =-\int_{-\infty}^{0}F\left(  t\right)  \exp\left\{
itx\right\}  dt.\label{inv F-}%
\end{equation}

The inverse to $\left(  \ref{inv F+}\right)  $ and $\left(  \ref{inv F-}%
\right)  $ transforms may be written as
\begin{equation}
\frac{1}{2\pi}\int_{-\infty}^{\infty}f_{+}\left(  x\right)  \exp\left\{
-itx\right\}  dx=\left\{
\begin{array}
[c]{ccc}%
F\left(  t\right)  ; & \text{if} & t>0,\\
0; & \text{if} & t<0
\end{array}
\right. \label{inverce}%
\end{equation}
and%
\begin{equation}
\frac{1}{2\pi}\int_{-\infty}^{\infty}f_{-}\left(  x\right)  \exp\left\{
-itx\right\}  dx=\left\{
\begin{array}
[c]{ccc}%
0; & \text{if} & t>0,\\
-F\left(  t\right)  ; & \text{if} & t<0.
\end{array}
\right. \label{inverce-}%
\end{equation}

Similarly one may write
\begin{equation}
F\left(  t\right)  =F_{+}\left(  t+i\varepsilon\right)  -F_{-}\left(
t-i\varepsilon\right) \label{vla-i}%
\end{equation}
where
\begin{equation}
F_{+}\left(  t\right)  =\frac{1}{2\pi}\int_{-\infty}^{0}f\left(  x\right)
\exp\left\{  -ixt\right\}  dx;\label{F+}%
\end{equation}
and
\begin{equation}
F_{-}\left(  t\right)  =-\frac{1}{2\pi}\int_{0}^{\infty}f\left(  x\right)
\exp\left\{  -ixt\right\}  dx\label{F-}%
\end{equation}

Representations $\left(  \ref{vla-i}\right)  $ and $\left(  \ref{inv-AP}%
\right)  $ are not unique. For instance, one may subtract the same entire
function $\Xi\left(  x\right)  $ from $f_{+}\left(  x\right)  $ and
$f_{-}\left(  x\right)  $ \cite{vladimirov,bremermann}. In particular, the
analytical properties of $f_{\pm}\left(  x\right)  $ are not changed, if one
chooses $\Xi\left(  x\right)  =$ $\int_{0}^{\xi}F\left(  t\right)
\exp\left\{  ixt\right\}  dt$. It means that decomposition of the domain of
integration is done at arbitrary finite $t=\xi$. The similar statement is
evidently true for $F_{\pm}\left(  t\right)  $.

It is evident\textrm{ }that if $\varepsilon$ is finite, $F\left(  t\right)  $
in $\left(  \ref{vla-i}\right)  $ and $f\left(  x\right)  $ in $\left(
\ref{inv-AP}\right)  $ are regular functions of their arguments for all real
$t$ and $x$. They must be treated as distributions only when $\varepsilon
\rightarrow0$.

\section{ Asymptotic Taylor series for distributions}

From $\left(  \ref{inv F+}\right)  $ it follows that%
\begin{equation}
e^{-i\tau x}f_{+}\left(  x\right)  =\int_{-\tau}^{\infty}F\left(
\tau+t\right)  e^{itx}dt
\end{equation}
so taking into account $\left(  \ref{vla-i}\right)  $ one may write%
\begin{equation}
e^{-i\tau x}f_{+}\left(  x\right)  =\int_{-\tau}^{\infty}\left[  F_{+}\left(
\tau+t+i\varepsilon\right)  -F_{-}\left(  \tau+t-i\varepsilon\right)  \right]
e^{itx}dt.\label{f+ Int}%
\end{equation}
Since $F_{\pm}\left(  t\right)  $ are analytical functions in upper/lower
complex half-plane $t$, then for any finite $\varepsilon$ and real $t$ and
$\tau$ they may be expanded in a Taylor series%

\begin{equation}
F_{\pm}\left(  \tau+t\pm i\varepsilon\right)  =\sum_{n=0}^{\infty}\frac
{F_{\pm}^{\left(  n\right)  }\left(  \tau\pm i\varepsilon\right)  }{n!}%
t^{n}\label{Taylor+/-}%
\end{equation}
that we write simply as%
\begin{equation}
F\left(  t+\tau\right)  =\sum_{n=0}^{\infty}\frac{t^{n}}{n!}F^{\left(
n\right)  }\left(  \tau\right) \label{T-or}%
\end{equation}
where%
\begin{equation}
F^{\left(  n\right)  }\left(  \tau\right)  \equiv F_{+}^{\left(  n\right)
}\left(  \tau+i\varepsilon\right)  -F_{-}^{\left(  n\right)  }\left(
\tau-i\varepsilon\right)  .
\end{equation}

If distribution $f\left(  x\right)  $ belongs to linear space $E^{\prime}$,
i.e. it is defined as functional $\left\langle T\left(  x\right)  ,\phi\left(
x\right)  \right\rangle $ on infinitely differentiable test functions
$\phi\left(  x\right)  \in E$ with arbitrary support\footnote{In \cite{zem} it
is only such distributions that are called \textit{generalized functions}.},
then its Fourier transform $%
%TCIMACRO{\tciFourier}%
%BeginExpansion
\mathcal{F}%
%EndExpansion
\left[  f,t\right]  =F\left(  t\right)  $ is an entire function
\cite{gel-shil,bremermann}, so Taylor series for $F\left(  t+\tau\right)  $
converges in the entire complex plane $t$. It is evident that even for
exponentially increasing $F_{\pm}^{\left(  n\right)  }\left(  \tau\right)  $
with $n$, the radii of convergence
\begin{equation}
R_{\pm}\equiv\left(  \lim_{n\rightarrow\infty}\left\vert \frac{F_{\pm
}^{\left(  n\right)  }\left(  \tau\pm i\varepsilon\right)  }{n!}\right\vert
^{-\frac{1}{n}}\right) \label{radius}%
\end{equation}
are infinite. Tempered distributions $F_{\pm}^{\left(  n\right)  }\left(
\tau\right)  $ may show quicker increase with $n$ and then radii may appear to
be finite. It may happen, when decreasing $F_{\pm}^{\left(  n\right)  }\left(
\tau\right)  $ with $\tau$ is fast enough. Indeed, simple qualitative analysis
shows that, e.g.
\begin{equation}
F_{-}^{\left(  n\right)  }\left(  \tau\right)  =-\frac{1}{2\pi}\int
_{0}^{\infty}f\left(  x\right)  x^{n}\exp\left\{  -ix\tau+n\ln x-i\pi
n/2\right\}  dx,
\end{equation}
have the quickest increase with $n$ when the main contribution in the integral
comes from large $x$. In this case, however, term $-ix\tau$ introduces
oscillation, thus providing a decrease with $\tau$. On the other hand, an
increasing $F_{+}^{\left(  n\right)  }\left(  \tau\right)  $ with $\tau$ is
the largest when the principal contribution comes from singularity (if any) of
$f\left(  x\right)  $ which is located in the complex upper half-plane $x,$
that, in turn, brings an oscillation with increasing $n$ and reduces
contribution from this area. It brings us to a suggestion that
\begin{equation}
F_{\pm}^{\left(  n+1\right)  }\left(  \tau\pm i\varepsilon\right)  =o\left(
F_{\pm}^{\left(  n\right)  }\left(  \tau\pm i\varepsilon\right)  \right)
\label{asy-cond}%
\end{equation}
should be considered. Condition $\left(  \ref{asy-cond}\right)  $ means that
$F_{\pm}^{\left(  n\right)  }\left(  \tau\pm i\varepsilon\right)  $
constitutes an asymptotic sequence and Taylor series $\left(  \ref{Taylor+/-}%
\right)  $ must be treated as asymptotic, i.e. $\left(  \ref{Taylor+/-}%
\right)  $ must be interpreted as%
\begin{equation}
F_{\pm}\left(  \tau+t\pm i\varepsilon\right)  =\sum_{n=0}^{N}\frac{F_{\pm
}^{\left(  n\right)  }\left(  \tau\pm i\varepsilon\right)  }{n!}t^{n}+o\left(
F_{\pm}^{\left(  N\right)  }\left(  \tau\pm i\varepsilon\right)  \right)
.\label{T_A}%
\end{equation}

We can not impose a condition similar to $\left(  \ref{asy-cond}\right)  $ on
$F^{\left(  n\right)  }\left(  \tau\right)  $ directly, because $F^{\left(
n+1\right)  }\left(  \tau\right)  =o\left(  F^{\left(  n\right)  }\left(
\tau\right)  \right)  $ simply means $F^{\left(  n+1\right)  }\left(
\tau\right)  /F^{\left(  n\right)  }\left(  \tau\right)  \rightarrow0$ for
$\tau\rightarrow\infty$ (or $\tau\rightarrow-\infty)$. However, neither
product nor ratio are defined for arbitrary distribution. Fortunately even
such an approach is not hopeless, because many distributions may be treated as
\textit{regular} in some area (see e.g. \cite{vladimirov}). In particular, if
$F^{\left(  n\right)  }\left(  \tau\right)  $ is regular, there exists
function $\psi^{\left(  n\right)  }\left(  t\right)  =\frac{\partial^{n}%
}{\partial t^{n}}\psi\left(  t\right)  $ with $n=0,..N$ such that
$\psi^{\left(  N\right)  }\left(  t\right)  $ is continuous and%
\begin{equation}
\left\langle F^{\left(  n\right)  }\left(  t\right)  ,\phi_{\Delta}\left(
t\right)  \right\rangle =\left\langle \psi^{\left(  n\right)  }\left(
t\right)  ,\phi_{\Delta_{\tau}}\left(  t\right)  \right\rangle \label{def}%
\end{equation}
for all test functions $\phi_{\Delta_{\tau}}\left(  t\right)  \in S$ with the
support located in the interval $\Delta_{\tau}=\left(  \tau-\Delta,\tau
+\Delta\right)  $, where positive parameter $\Delta$ may be chosen as
arbitrary small, but must remain finite. Functions $\psi^{\left(  n\right)
}\left(  t\right)  $ are unique and may be treated as asymptotics of
$F^{\left(  n\right)  }\left(  t\right)  $.

Since tempered distribution $F\left(  \tau\right)  $ is defined by conditions
$\left(  \ref{cond_t}\right)  $ and $\left(  \ref{cond_n}\right)  $, it is
natural to assume that for\footnote{Proceeding to limit $\tau\rightarrow
-\infty$ is very similar, but constants may differ.} $\tau\rightarrow\infty$
\begin{equation}
\psi\left(  \tau\right)  \sim C\tau^{-\gamma}\exp\left\{  -\lambda\tau
^{\alpha}\right\}
\end{equation}
where constants $C$, $\gamma$ and $\alpha$ are arbitrary and $\lambda$ is
nonnegative. Asymptotic behavior of $\exp\left\{  -\lambda\tau^{\alpha
}\right\}  $ for $\alpha<0$ is trivial. For $\alpha>0$ for $\tau
\rightarrow\infty$ we get
\begin{equation}
\psi^{\left(  n\right)  }\left(  \tau\right)  \sim\left(  -\lambda\alpha
\tau^{\alpha-1}\right)  ^{n}\psi\left(  \tau\right)
\end{equation}
so increasing $F^{\left(  n\right)  }\left(  \tau\right)  $ with $n$ remains
exponential and $R$ is infinite. However, for $\alpha=0$ we obtain
\begin{equation}
\psi^{\left(  n\right)  }\left(  \tau\right)  \sim\frac{\partial^{n}}%
{\partial\tau^{n}}\tau^{-\gamma}=\left(  -1\right)  ^{n}\frac{\Gamma\left(
n+\gamma\right)  }{\Gamma\left(  \gamma\right)  }\tau^{-\gamma-n}%
\end{equation}
and%

\begin{equation}
R=\lim_{n\rightarrow\infty}\left\vert \frac{\psi^{\left(  n\right)  }\left(
\tau\right)  }{n!}\right\vert ^{-\frac{1}{n}}=\tau.
\end{equation}
In this case Taylor series do not converge for $\left\vert t\right\vert
>\left\vert \tau\right\vert $, but it is not too restrictive because we intend
to increase $\left\vert \tau\right\vert $ infinitely.

Now, taking into account%
\begin{equation}
\frac{1}{2\pi}\int_{-\tau}^{\infty}t^{n}e^{itx}dt=\left\{
\begin{array}
[c]{ccc}%
i^{-n}\delta^{\left(  n\right)  }\left(  x\right)  ; & \text{if} &
\tau\rightarrow\infty,\\
0; & \text{if} & \tau\rightarrow-\infty.
\end{array}
\right. \label{bry}%
\end{equation}
(see e.g. \cite{brych}), in $\left(  \ref{f+ Int}\right)  $ we may confine
ourselves to area $\left\vert t/\tau\right\vert <1$, so
\begin{equation}
e^{-i\tau x}f_{+}\left(  x\right)  =\sum_{n=0}^{\infty}\frac{F^{\left(
n\right)  }\left(  \tau\right)  }{n!}\frac{1}{2\pi}\int_{-\tau}^{\tau}%
t^{n}e^{itx}dt
\end{equation}
and we may write
\begin{equation}
e^{-i\tau x}f_{+}\left(  x\right)  =\left\{
\begin{array}
[c]{ccc}%
2\pi\sum_{n=0}^{\infty}\frac{F^{\left(  n\right)  }\left(  \tau\right)  }%
{n!}i^{-n}\delta^{\left(  n\right)  }\left(  x\right)  ; & \text{if} &
\tau\rightarrow\infty,\\
0; & \text{if} & \tau\rightarrow-\infty.
\end{array}
\right. \label{ba+}%
\end{equation}

In the same way, from%
\begin{equation}
e^{-i\tau x}f_{-}\left(  x\right)  =-\int_{-\infty}^{-\tau}F\left(
\tau+t\right)  e^{itx}dt
\end{equation}
we find%
\begin{equation}
e^{-i\tau x}f_{-}\left(  x\right)  =\left\{
\begin{array}
[c]{ccc}%
0; & \text{if} & \tau\rightarrow\infty,\\
-2\pi\sum_{n=0}^{\infty}\frac{F^{\left(  n\right)  }\left(  \tau\right)  }%
{n!}i^{-n}\delta^{\left(  n\right)  }\left(  x\right)  ; & \text{if} &
\tau\rightarrow-\infty.
\end{array}
\right.  \label{ba --}%
\end{equation}
From $\left(  \ref{ba --}\right)  $,$\left(  \ref{ba+}\right)  $ and $\left(
\ref{inv-AP}\right)  $ we also get for $\tau\rightarrow\pm\infty$%
\begin{equation}
e^{-i\tau x}f\left(  x\right)  =2\pi\sum_{n=0}^{\infty}\frac{F^{\left(
n\right)  }\left(  \tau\right)  }{n!}i^{-n}\delta^{\left(  n\right)  }\left(
x\right)  .\label{ba}%
\end{equation}

To finalize this section we wish to suggest a simple interpretation of the
condition $\left(  \ref{asy-cond}\right)  $, namely what it means for
$f\left(  x\right)  $. Since $\left(  \ref{asy-cond}\right)  $ is satisfied
for $t^{\prime}>\tau$, we can write an asymptotic series%
\begin{equation}
F_{\pm}\left(  t+t^{\prime}\pm i\varepsilon\right)  \exp\left\{  it^{\prime
}x\right\}  =\sum_{n=0}^{\infty}\frac{1}{n!}F_{\pm}^{\left(  n\right)
}\left(  t^{\prime}\pm i\varepsilon\right)  \exp\left\{  it^{\prime}x\right\}
t^{n}.
\end{equation}
The asymptotic series allows termwise integration \cite{Erdelyi}, so%
\begin{equation}
\int_{\tau}^{\infty}F_{\pm}\left(  t+t^{\prime}\pm i\varepsilon\right)
\exp\left\{  it^{\prime}x\right\}  dt^{\prime}=\sum_{n=0}^{\infty}\frac{1}%
{n!}t^{n}s_{\pm}^{\left[  n\right]  }\left(  x,\tau\right)
\end{equation}
with%
\begin{equation}
s_{\pm}^{\left[  n\right]  }\left(  x,\tau\right)  =\int_{\tau}^{\infty}%
F_{\pm}^{\left(  n\right)  }\left(  t^{\prime}\pm i\varepsilon\right)
\exp\left\{  it^{\prime}x\right\}  dt^{\prime}%
\end{equation}
is an asymptotic series too and the condition $\left(  \ref{asy-cond}\right)
$ transforms into%
\begin{equation}
s_{\pm}^{\left[  n\right]  }\left(  x,\tau\right)  =o\left(  s_{\pm}^{\left[
n\right]  }\left(  x,\tau\right)  \right)  .
\end{equation}
One may rewrite $\left(  \ref{inv F+}\right)  $ as%
\begin{equation}
f_{+}\left(  x\right)  =\int_{\tau}^{\infty}\left[  F_{+}\left(
t+i\varepsilon\right)  -F_{-}\left(  t-i\varepsilon\right)  \right]
\exp\left\{  itx\right\}  dt+f_{+}^{reg}\left(  x\right)
\end{equation}
where
\begin{equation}
f_{+}^{reg}\left(  x\right)  =\int_{0}^{\tau}\left[  F_{+}\left(
t+i\varepsilon\right)  -F_{-}\left(  t-i\varepsilon\right)  \right]
\exp\left\{  itx\right\}  dt.\label{reg}%
\end{equation}
Since integration in $\left(  \ref{reg}\right)  $ is done over the finite
interval, $f_{+}^{reg}\left(  x\right)  $ is regular part of $f_{+}\left(
x\right)  $, so that $s_{\pm}^{\left[  0\right]  }\left(  x,\tau\right)  $
reproduces singularity of $f_{+}\left(  x\right)  $ at $x=0$, if any. At
$x\neq0$ oscillations remove singularity.

It is clear that
\begin{equation}
s_{\pm}^{\left[  n\right]  }\left(  x,\tau\right)  =-F_{\pm}^{\left(
n-1\right)  }\left(  \tau\pm i\varepsilon\right)  \exp\left\{  i\tau
x\right\}  -ixs_{\pm}^{\left[  n-1\right]  }\left(  x,\tau\right)
.\label{part}%
\end{equation}

If we assume $s_{\pm}^{\left[  n\right]  }\left(  x,\tau\right)  \sim o\left(
F_{\pm}^{\left(  n-1\right)  }\left(  \tau\pm i\varepsilon\right)  \right)  $
$\left(  \ref{part}\right)  $ immediately leads to $s_{\pm}^{\left[  n\right]
}\left(  x,\tau\right)  \sim\frac{1}{ix}F_{\pm}^{\left(  n\right)  }\left(
\tau\pm i\varepsilon\right)  $ and consequently to $F_{\pm}^{\left(  n\right)
}\left(  \tau\pm i\varepsilon\right)  /x\sim o\left(  F_{\pm}^{\left(
n-1\right)  }\left(  \tau\pm i\varepsilon\right)  \right)  $; then, since
$\tau$ is fixed, this condition will be broken for small enough $x$. However,
if $F_{\pm}^{\left(  n-1\right)  }\left(  \tau\pm i\varepsilon\right)  \sim
s_{\pm}^{\left[  n\right]  }\left(  x,\tau\right)  $ or $F_{\pm}^{\left(
n-1\right)  }\left(  \tau\pm i\varepsilon\right)  \sim o\left(  s_{\pm
}^{\left[  n\right]  }\left(  x,\tau\right)  \right)  $ then $s_{\pm}^{\left[
n\right]  }\left(  x,\tau\right)  \sim ixs_{\pm}^{\left[  n-1\right]  }\left(
x,\tau\right)  $ or
\begin{equation}
s_{\pm}^{\left[  n\right]  }\left(  x,\tau\right)  \simeq\left(  -ix\right)
^{n}s_{\pm}^{\left[  0\right]  }\left(  x,\tau\right)  .
\end{equation}
Hence, condition $\left(  \ref{asy-cond}\right)  $ looks quite natural, since
it means a reduction of singularity degree when $x^{n-1}f_{+}\left(  x\right)
$ it is multiplied by $x$.

\section{Series for distributions with $\tau$-dependent regularization
parameter}

Until now we dealt with the distributions defined by $\left(  \ref{inv-AP}%
\right)  $. It implies that $\left\vert \tau\right\vert $ may be taken
arbitrary large, but transition to the limit $\tau\rightarrow\pm\infty$ must
be carried out only after taking the limit $\varepsilon\rightarrow0$. Here we
consider another way of taking this limit, namely
\begin{equation}
e^{-i\tau x}f_{+}\left(  x+i\frac{\nu}{\tau}\right)  =e^{-i\tau x}f_{+}\left(
x+i\frac{\nu}{\tau}\right)  =\int_{-\tau}^{\infty}F\left(  \tau+t\right)
e^{it\left(  x+i\frac{\nu}{\tau}\right)  }dt;\quad\nu>0.
\end{equation}
Entire function $\exp\left\{  it\left(  x+i\frac{\nu}{\tau}\right)  \right\}
$ may be expanded in a Taylor series for arbitrary $t\nu/\tau$, but to
preserve the uniform convergence of the integral we left infinitesimal
parameter $\varepsilon$ in the exponent. In this case the interchange of
summations and integration order is legal and we may write
\begin{equation}
e^{-i\tau x}f_{+}\left(  x+i\frac{\nu}{\tau}\right)  =\sum_{m=0}^{\infty}%
\frac{\left(  -\frac{\nu}{\tau}+\varepsilon\right)  ^{m}}{m!}\int_{-\tau
}^{\infty}F\left(  \tau+t\right)  t^{m}e^{it\left(  x+i\varepsilon\right)
}dt.
\end{equation}
After taking the limit $\varepsilon\rightarrow0$ we obtain%
\begin{equation}
e^{-i\tau x}f_{+}\left(  x+i\frac{\nu}{\tau}\right)  =\sum_{m=0}^{\infty}%
\frac{\left(  -\frac{\nu}{\tau}\right)  ^{m}}{m!}\int_{-\tau}^{\infty}F\left(
\tau+t\right)  t^{m}e^{it\left(  x+i0\right)  }dt
\end{equation}
that leads to%
\begin{equation}
e^{-i\tau x}f_{+}\left(  x+i\frac{\nu}{\tau}\right)  =\int_{-\tau}^{\tau}%
\sum_{m=0}^{\infty}\frac{\left(  -\frac{\nu}{\tau}\right)  ^{m}}{m!}t^{m}%
\sum_{n=0}^{\infty}\frac{F^{\left(  n\right)  }\left(  \tau\right)  }{n!}%
t^{n}e^{it\left(  x+i0\right)  }dt.
\end{equation}
Since $t$ in Taylor series is located in the convergence area, we may write%
\begin{equation}
\sum_{m=0}^{\infty}\frac{\left(  -\frac{\nu}{\tau}\right)  ^{m}}{m!}t^{m}%
\sum_{n=0}^{\infty}\frac{F^{\left(  n\right)  }\left(  \tau\right)  }{n!}%
t^{n}=2\pi\sum_{m=0}^{\infty}\left(  it\right)  ^{n}C_{n}\left(  \nu
,\tau\right)
\end{equation}
with%
\begin{equation}
C_{n}\left(  \tau,\nu/\tau\right)  =\frac{i^{-n}}{2\pi}\sum_{n=k}^{\infty
}\left(  \frac{F^{\left(  n-k\right)  }\left(  \tau\right)  }{\left(
n-k\right)  !}\frac{\left(  -\frac{\nu}{\tau}\right)  ^{k}}{k!}\right)
\end{equation}
and we finally get%

\begin{equation}
e^{-i\tau x}f_{+}\left(  x+i\frac{\nu}{\tau}\right)  =\sum_{n=0}^{\infty}%
C_{n}\left(  \tau,\nu/\tau\right)  \delta^{\left(  n+m\right)  }\left(
x\right)  .\label{exp-nu}%
\end{equation}

In the same way, from%
\begin{equation}
e^{-i\tau x}f_{-}\left(  x-i\frac{\nu}{\tau}\right)  =-\int_{-\infty}^{-\tau
}F\left(  \tau+t\right)  e^{it\left(  x-i\frac{\nu}{\tau}\right)  }dt
\end{equation}
we find%
\begin{equation}
e^{-i\tau x}f_{-}\left(  x\right)  =\left\{
\begin{array}
[c]{ccc}%
0; & \text{if} & \tau\rightarrow\infty,\\
-\sum_{n=0}^{\infty}C_{n}\left(  \tau,\nu/\left\vert \tau\right\vert \right)
\delta^{\left(  n\right)  }\left(  x\right)  ; & \text{if} & \tau
\rightarrow-\infty.
\end{array}
\right.
\end{equation}
It easy to check that expression $\left(  \ref{exp-nu}\right)  $ may be
rewritten as%
\begin{equation}
e^{-i\tau x}\left[  f_{+}\left(  x+i\frac{\nu}{\tau}\right)  -f_{-}\left(
x-i\frac{\nu}{\tau}\right)  \right]  =\sum_{n=0}^{\infty}C_{n}\left(  \tau
,\nu/\left\vert \tau\right\vert \right)  \delta^{\left(  n\right)  }\left(
x\right) \label{exp-nu 2}%
\end{equation}
with%
\begin{equation}
C_{n}\left(  \tau\right)  =2\pi i^{-n}\sum_{m=0}^{n}\frac{F^{\left(
n-m\right)  }\left(  \tau\right)  }{\left(  n-m\right)  !}\frac{\left(
-\nu/\left\vert \tau\right\vert \right)  ^{m}}{m!}.
\end{equation}
It is clear that with $\nu\rightarrow0$ we return to $\left(  \ref{ba}\right)
$.

\section{Asymptotic expansion for power type distribution}

Consider as an example expansion for $\exp\left\{  -ix\tau\right\}  $ $\left(
x+i\nu/\tau\right)  ^{\lambda}$. With\footnote{An Abel-Poisson regularization
is implied.}%
\begin{equation}
\left(  x+i\nu/\tau\right)  ^{\lambda}=\sum_{n=0}^{\infty}\frac{\Gamma\left(
-\lambda+n\right)  }{n!\Gamma\left(  -\lambda\right)  }\left(  -i\nu
/\tau\right)  ^{n}\left(  x+i\varepsilon\operatorname*{signum}\left(
\tau\right)  \vspace{6pt}\right)  ^{\lambda-n}%
\end{equation}
and taking into account \cite{brych-prud}8.8$\left(  677\right)  $ one may
obtain%
\begin{align}
F\left(  t\right)   & =\frac{1}{2\pi}\int_{-\infty}^{\infty}\left(
x+i\frac{\nu}{\tau}\right)  ^{\lambda}e^{-ixt}\,dx\nonumber\\
& =\left\{
\begin{array}
[c]{ccc}%
\frac{e^{i\pi\frac{\lambda}{2}}}{\Gamma\left(  -\lambda\right)  }\sum
_{n=0}^{\infty}\frac{\left(  \frac{\nu}{\tau}\right)  ^{n}t_{+}^{n-\lambda-1}%
}{n!}; & \text{if} & \tau>0,\\
\frac{e^{-i\pi\frac{\lambda}{2}}}{\Gamma\left(  -\lambda\right)  }\sum
_{n=0}^{\infty}\frac{\left(  -\frac{\nu}{\left\vert \tau\right\vert }\right)
^{n}t_{-}^{n-\lambda-1}}{n!}; & \text{if} & \tau<0.
\end{array}
\right.  \allowbreak
\end{align}
It is clear that for $\tau\rightarrow\pm\infty$ and $\left\vert t\right\vert
<\left\vert \tau\right\vert $ we get$\allowbreak$%
\begin{equation}
F\left(  t+\tau\right)  =\left\{
\begin{array}
[c]{ccc}%
e^{\nu+i\pi\frac{\lambda}{2}}\frac{\left(  t+\tau\right)  ^{-\lambda-1}%
}{\Gamma\left(  -\lambda\right)  }e^{\frac{\nu}{\tau}t} & \text{if} &
\tau\rightarrow\infty,\\
e^{-\nu-i\pi\frac{\lambda}{2}}\frac{\left(  \left\vert \tau\right\vert
-t\right)  ^{-\lambda-1}}{\Gamma\left(  -\lambda\right)  }e^{\frac{\nu
}{\left\vert \tau\right\vert }t} & \text{if} & \tau\rightarrow-\infty.
\end{array}
\right.  \allowbreak
\end{equation}

In the area $\left\vert t\right\vert <\tau$ one may also write%

\begin{equation}
\frac{\left(  \left\vert \tau\right\vert \pm t\right)  ^{-\lambda-1}}%
{\Gamma\left(  -\lambda\right)  }=\bigskip\sum_{n=0}^{\infty}\frac{1}%
{n!\Gamma\left(  -\lambda-n\right)  }\left(  \pm t\right)  ^{n}\left\vert
\tau\right\vert ^{-\lambda-1-n}%
\end{equation}
so%
\begin{equation}
F\left(  t+\tau\right)  =\left\{
\begin{array}
[c]{ccc}%
e^{\nu+i\pi\frac{\lambda}{2}}\sum_{n,m=0}^{\infty}\frac{\tau^{-\lambda
-1-n}t^{n}}{n!\Gamma\left(  -\lambda-n\right)  }\frac{\nu^{m}\left\vert
\tau\right\vert ^{-m}t^{m}}{m!}; & \text{if} & \tau\rightarrow\infty,\\
e^{-\nu-i\pi\frac{\lambda}{2}}\sum_{n,m=0}^{\infty}\frac{\left(  -1\right)
^{n}\left\vert \tau\right\vert ^{-\lambda-1-n}t^{n}}{n!\Gamma\left(
-\lambda-n\right)  }\frac{\nu^{m}\left\vert \tau\right\vert ^{-m}t^{m}}{m!}; &
\text{if} & \tau\rightarrow-\infty,
\end{array}
\right.  \allowbreak
\end{equation}
or%
\begin{equation}
\allowbreak F\left(  t+\tau\right)  =\left\{
\begin{array}
[c]{ccc}%
e^{\nu+i\pi\frac{\lambda}{2}}\sum_{n=0}^{\infty}p_{n}^{\lambda}\nu
\tau^{-\lambda-1-n}t^{n}; & \text{if} & \tau\rightarrow\infty,\\
e^{-\nu-i\pi\frac{\lambda}{2}}\sum_{n=0}^{\infty}p_{n}^{\lambda}\left(
-\nu\right)  \left\vert \tau\right\vert ^{-\lambda-1-n}\left(  -t\right)
^{n}; & \text{if} & \tau\rightarrow-\infty,
\end{array}
\right.  \allowbreak
\end{equation}
where%
\begin{equation}
p_{n}^{\lambda}\left(  \nu\right)  \equiv\sum_{m=0}^{n}\frac{\nu^{m}%
}{m!\left(  n-m\right)  !\Gamma\left(  m-\lambda-n\right)  }%
\end{equation}
are polynomials\footnote{Up to a constant $p_{n}^{\lambda}\left(  \nu\right)
$ are a finite Kummer series (see \cite{bateman} 6.1).} of degree $n$. So
having collected everything we obtain%
\begin{equation}
\allowbreak e^{-ix\tau}\left(  x+i\frac{\nu}{\tau}\right)  ^{\lambda}=\left\{
\begin{array}
[c]{ccc}%
e^{\nu+i\pi\frac{\lambda}{2}}\sum_{n=0}^{\infty}p_{n}^{\lambda}\left(
\nu\right)  \tau^{-\lambda-1-n}i^{-n}\delta^{(n)}\left(  x\right)  ; &
\text{if} & \tau\rightarrow\infty\\
e^{-\nu-i\pi\frac{\lambda}{2}}\sum_{n=0}^{\infty}p_{n}^{\lambda}\left(
-\nu\right)  \left\vert \tau\right\vert ^{-\lambda-1-n}i^{n}\delta
^{(n)}\left(  x\right)  ; & \text{if} & \tau\rightarrow-\infty.
\end{array}
\right.
\end{equation}
It goes without saying that for $\tau\rightarrow\pm\infty$%
\begin{equation}
\allowbreak e^{\pm ix\tau}\left(  x\pm i\frac{\nu}{\tau}\right)  ^{\lambda}=0
\end{equation}
$\allowbreak$

\section{Conclusion}

A simple method is suggested to compute a series for $e^{-i\tau x}f\left(
x\right)  $ for $\tau\rightarrow\infty$. It is shown that $n$-th term in the
series is proportional to $n$-th derivative of the Fourier transform $f\left(
x\right)  $, and namely to $F^{\left(  n\right)  }\left(  \tau\right)  $. If
$f\left(  x\right)  \in E^{\prime}$, i.e. defined as functional $\left\langle
f\left(  x\right)  ,\phi\left(  x\right)  \right\rangle $ on infinitely
differentiable test functions $\phi\left(  x\right)  \in E$ with arbitrary
support, then $F\left(  t+\tau\right)  $ is an entire function and Taylor
series converges in the whole complex plane $t$. A more interesting case
appears when we consider tempered distributions $f\left(  x\right)  $, such
that $F\left(  t+\tau\right)  $ allows asymptotic expansions, i.e., roughly
speaking, $F_{\pm}^{\left(  n+1\right)  }\left(  \tau\right)  =O\left(
F_{\pm}^{\left(  n\right)  }\left(  \tau\right)  /\tau\right)  $ for
$\tau\rightarrow\pm\infty$. In such case the Taylor series for $F\left(
t+\tau\right)  $ converges only in the bounded area $\left\vert t\right\vert
<\left\vert \tau\right\vert $, but an asymptotic expansion still exists.

As it is shown in Appendix, if distribution $F\left(  t\right)  $\ allows the
Mellin transform and obeys conditions specified there, the asymptotic series
$\left(  \ref{ba+}\right)  $\ and $\left(  \ref{ba --}\right)  $\ may be
obtained with the asymptotic expansion for a particular type of power-type
distributions $e^{i\tau x}\left(  x\pm i\varepsilon\right)  ^{-s}$ computed in
\cite{brych,brych-prud}.

\section{Appendix. Modified Mellin transform}

Comprehensive study of Mellin transform%
\begin{equation}
m\left(  s\right)  =\int_{0}^{\infty}t^{s-1}F\left(  t\right)  dt\label{Mell}%
\end{equation}
for distributions was undertaken in \cite{zem}. It is shown, in particular,
that if $m\left(  s\right)  $ is represented by $\left(  \ref{Mell}\right)
$\ for $s\in\Lambda$, then $m\left(  s\right)  $\ is regular in $\Lambda$.
Moreover, if $m\left(  s\right)  $ is regular in strip $a<\operatorname{Re}%
s<b$ and
\begin{equation}
\left\vert m\left(  s\right)  \right\vert <C_{0}\left\vert s\right\vert
^{-2},\quad C_{0}=constant,\label{restr-m(s)}%
\end{equation}
then the inverse Mellin transform may be written as%
\begin{equation}
F\left(  t\right)  =\frac{1}{2\pi i}\int_{\sigma-i\infty}^{\sigma+i\infty
}m\left(  s\right)  t^{-s}ds\label{inverce-M}%
\end{equation}
where $a<\sigma<b$.

Integration path in $\left(  \ref{Mell}\right)  $ includes the area of
positive $t$ only, hence $m\left(  s\right)  $ carries no information about
$F\left(  t\right)  $ behavior for $t<0$. Expression $\left(  \ref{inverce-M}%
\right)  $ may allow to compute $F\left(  t\right)  $ for some $t<0$ as an
analytical extension, however, it may appear unacceptable for $F\left(
t\right)  $ being nonanalytic for $t<0$.

To represent the distribution $F\left(  t\right)  $ with Mellin transform in
the whole definitional domain of $F\left(  t\right)  $, we divide this domain
into positive and negative parts. Mellin transform of $F\left(  t\right)  $
for $t>0$ is
\begin{equation}
m_{+}\left(  s\right)  =\int_{0}^{\infty}t^{s-1}F\left(  t\right)
dt=\int_{-\infty}^{\infty}t_{+}^{s-1}F\left(  t\right)  dt\label{Mell-M}%
\end{equation}
and, in fact, it coincides with $\left(  \ref{Mell}\right)  $. For $t<0$ we
relate $F\left(  t\right)  $ to
\begin{equation}
m_{-}\left(  s\right)  =\int_{0}^{\infty}t^{s-1}F\left(  -t\right)
dt=\int_{-\infty}^{0}\left\vert t\right\vert ^{s-1}F\left(  t\right)
dt=\int_{-\infty}^{\infty}t_{-}^{s-1}F\left(  t\right)  dt\label{Mell-M-}%
\end{equation}
where by the definition (see e.g. \cite{gel-shil})%
\begin{equation}
t_{+}^{\lambda}\triangleq\left\{
\begin{array}
[c]{ccc}%
t^{\lambda}; & \text{if} & t>0\\
0; & \text{if} & t<0
\end{array}
\right.  ;\quad t_{-}^{\lambda}\triangleq\left\{
\begin{array}
[c]{ccc}%
0; & \text{if} & t>0\\
\left\vert t\right\vert ^{\lambda}; & \text{if} & t<0
\end{array}
\right.
\end{equation}

If we assume that $m_{\pm}\left(  s\right)  $\ are regular in the strips
$a_{\pm}<\operatorname{Re}s<b_{\pm}$, then inverse Mellin transforms acquire
the form%
\begin{equation}
\frac{1}{2\pi i}\int_{\sigma_{+}-i\infty}^{\sigma_{+}+i\infty}m_{+}\left(
s\right)  t_{+}^{-s}ds=\left\{
\begin{array}
[c]{ccc}%
F\left(  t\right)  ; & \text{if} & t>0\\
0; & \text{if} & t<0
\end{array}
\right. \label{Mell-Inverce}%
\end{equation}
and%
\begin{equation}
\frac{1}{2\pi i}\int_{\sigma_{-}-i\infty}^{\sigma_{-}+i\infty}m_{-}\left(
s\right)  t_{-}^{-s}ds=\left\{
\begin{array}
[c]{ccc}%
0; & \text{if} & t>0\\
F\left(  t\right)  ; & \text{if} & t<0
\end{array}
\right.  .\label{Mell-Inverce-}%
\end{equation}
where $a_{\pm}<\sigma_{\pm}<b_{\pm}$.

If strips $a_{\pm}<\operatorname{Re}s<b_{\pm}$ have the common area
$a<\operatorname{Re}s<b $, where $a\geq a_{\pm}$ and $b\leq b_{\pm}$, then we
may write%
\begin{equation}
F\left(  t\right)  =\frac{1}{2\pi i}\int_{\sigma-i\infty}^{\sigma+i\infty
}\left(  m_{+}\left(  s\right)  t_{+}^{-s}+m_{-}\left(  s\right)  t_{-}%
^{-s}\right)  ds
\end{equation}
where $a<\sigma<b$.

Recall, that distributions $t_{\pm}^{-s}\ $are nonanalytic in $s$, namely,
they have simple poles for positive integer $s$%
\begin{equation}
t_{+}^{-s}=\frac{1}{s-n}\frac{\left(  -1\right)  ^{n}}{\left(  n-1\right)
!}\delta^{\left(  n-1\right)  }\left(  t\right)  +\xi_{n}\left(
t_{+},s\right) \label{pol}%
\end{equation}
where $\xi_{n}\left(  t_{+},s\right)  $ is the regular part of the Laurent
series \cite{gel-shil}. For our purpose it is convenient to modify Mellin
transform by changing $t_{\pm}^{-s}$ for distribution $\left(  x\pm
i\varepsilon\right)  ^{s-1}$ which depends on $s$ regularly. From
\cite{brych-prud}%
\begin{equation}
\int_{-\infty}^{\infty}t_{\pm}^{-s}\exp\left\{  itx\right\}  dt=\exp\left\{
\pm i\frac{\pi}{2}\left(  1-s\right)  \right\}  \Gamma\left(  1-s\right)
\left(  x\pm i\varepsilon\right)  ^{s-1}\label{b-p}%
\end{equation}
we see that it may be done by Fourier transform of $\left(  \ref{Mell-Inverce}%
\right)  $ and $\left(  \ref{Mell-Inverce-}\right)  $ that leads to
\begin{align}
& \frac{1}{2\pi i}\int_{\sigma_{+}-i\infty}^{\sigma_{+}+i\infty}m_{+}\left(
s\right)  \exp\left\{  i\frac{\pi}{2}\left(  1-s\right)  \right\}
\Gamma\left(  1-s\right)  \left(  x+i\varepsilon\right)  ^{s-1}%
ds\label{Mell-Four}\\
& =\int_{0}^{\infty}F\left(  t\right)  \exp\left\{  itx\right\}  dt\nonumber
\end{align}
and%
\begin{align}
& \frac{1}{2\pi i}\int_{\sigma_{-}-i\infty}^{\sigma_{-}+i\infty}m_{-}\left(
s\right)  \exp\left\{  -i\frac{\pi}{2}\left(  1-s\right)  \right\}
\Gamma\left(  1-s\right)  \left(  x-i\varepsilon\right)  ^{s-1}%
ds\label{Mell-Four-}\\
& =\int_{\infty}^{0}F\left(  t\right)  \exp\left\{  itx\right\}  dt\nonumber
\end{align}

Taking into account $\left(  \ref{inv F+}\right)  $ and $\left(
\ref{inv F-}\right)  $, and changing $s\rightarrow1-s$ in $\left(
\ref{Mell-Four}\right)  $, $\left(  \ref{Mell-Four-}\right)  $ we may write%
\begin{equation}
f_{+}\left(  x\right)  =\int_{0}^{\infty}F\left(  t\right)  \exp\left\{
itx\right\}  dt=\frac{1}{2\pi i}\int_{c_{+}-i\infty}^{c_{+}+i\infty}\mu
_{+}\left(  s\right)  \left(  x+i\varepsilon\right)  ^{-s}%
ds\label{modif-inverce}%
\end{equation}
and%
\begin{equation}
f_{-}\left(  x\right)  =-\int_{\infty}^{0}F\left(  t\right)  \exp\left\{
itx\right\}  dt=-\frac{1}{2\pi i}\int_{c_{-}-i\infty}^{c_{-}+i\infty}\mu
_{-}\left(  s\right)  \left(  x-i\varepsilon\right)  ^{-s}%
ds\label{modif-inverce-}%
\end{equation}
where%
\begin{equation}
\mu_{\pm}\left(  s\right)  =\exp\left\{  \pm is\pi/2\right\}  \Gamma\left(
s\right)  m_{\pm}\left(  1-s\right)  ;\quad c_{\pm}=1-\sigma_{\pm},
\end{equation}

The regularity of $m_{\pm}\left(  s\right)  $\ in strips $a_{\pm
}<\operatorname{Re}s<b_{\pm}$ causes the regularity of $\mu_{\pm}\left(
s\right)  $\ in $1-b_{\pm}<\operatorname{Re}s<1-a_{\pm}$\ except, possibly,
simple poles at negative integer values of $s=-n$, if such are located in the
considered strip. Such poles correspond to the ones in $\left(  \ref{pol}%
\right)  $. It is clear that if one shifts the path of integration in $\left(
\ref{modif-inverce}\right)  $ or $\left(  \ref{modif-inverce-}\right)  $ in
the area $1-b_{\pm}<\operatorname{Re}s<1-a_{\pm}$ with some pole crossing, a
result is obtained which differs from the original by the residues in such poles.

Since%
\begin{equation}
\lim_{\left\vert \operatorname{Im}s\right\vert \rightarrow\infty}\left\vert
\Gamma\left(  s\right)  \right\vert =\sqrt{2\pi}\exp\left\{  -\frac{\pi}%
{2}\left\vert \operatorname{Im}s\right\vert \right\}  \left\vert
\operatorname{Im}s\right\vert ^{\operatorname{Re}s-\frac{1}{2}}%
\end{equation}
(see \cite{bateman}1.1.18$\left(  6\right)  $)), we get from $\left(
\ref{restr-m(s)}\right)  $\ a restriction on $\mu_{\pm}\left(  s\right)  $\ in
strips $1-b_{\pm}<\operatorname{Re}s<1-a_{\pm}$\
\begin{equation}
\left\vert \mu_{\pm}\left(  s\right)  \right\vert <C_{0}\left\vert
1-s\right\vert ^{-2}\left\vert \operatorname{Im}s\right\vert
^{\operatorname{Re}s-\frac{1}{2}}\exp\left\{  -\pi\theta\left(  \pm
\operatorname{Im}s\right)  \left\vert \operatorname{Im}s\right\vert \right\}
\end{equation}
that for $\left\vert \operatorname{Im}s\right\vert \rightarrow\infty$\ leads to%

\begin{equation}
\left\vert \mu_{\pm}\left(  s\right)  \right\vert <C_{0}\left\vert
\operatorname{Im}s\right\vert ^{-\sigma_{\pm}-\frac{3}{2}}\exp\left\{
-\pi\theta\left(  \pm\operatorname{Im}s\right)  \left\vert \operatorname{Im}%
s\right\vert \right\}  .
\end{equation}
Fast decrease of $\left\vert \mu_{\pm}\left(  s\right)  \right\vert $\ for
$\operatorname{Im}s\rightarrow\pm\infty$\ compensates the increase of
$\left\vert \left(  x\pm i\varepsilon\right)  ^{-s}\right\vert =\left\vert
x\right\vert ^{-s}\exp\left\{  \pm\pi\operatorname{Im}s\right\}  $\ for
$x<0$\ and provides convergence of integral in $\left(  \ref{modif-inverce}%
\right)  $.

Expressions $\left(  \ref{modif-inverce}\right)  $ and $\left(
\ref{modif-inverce-}\right)  $ give convenient representations for the
functions $f_{\pm}\left(  x\right)  $ which are analytical in the upper/lower
half-plane of a complex variable $x$.

An application of inverse Mellin transform $\left(  \ref{Mell-Inverce}\right)
$ to $\left(  \ref{modif-inverce}\right)  $ relates $\mu_{\pm}\left(
s\right)  $ to $F\left(  t\right)  $%

\begin{equation}
\mu_{\pm}\left(  s\right)  =\exp\left\{  \pm i\frac{\pi}{2}s\right\}
\Gamma\left(  s\right)  \int_{-\infty}^{\infty}t_{\pm}^{-s}F\left(  t\right)
dt
\end{equation}
so with $\left(  \ref{b-p}\right)  $\ one may easily get%
\begin{equation}
\mu_{+}\left(  s\right)  =\frac{1}{1-e^{2i\pi s}}\int_{-\infty}^{\infty}%
f_{+}\left(  x\right)  \left(  x-i\varepsilon\right)  ^{s-1}%
dx\label{modif-Mell}%
\end{equation}
and%
\begin{equation}
\mu_{-}\left(  s\right)  =-\frac{1}{1-e^{2i\pi s}}\int_{-\infty}^{\infty}%
f_{-}\left(  x\right)  \left(  x+i\varepsilon\right)  ^{s-1}dx.
\end{equation}

With $\left(  \ref{modif-inverce}\right)  $ one may easily regain the series
$\left(  \ref{ba+}\right)  $. Indeed, with simple relation$\qquad$%
\begin{equation}
\frac{1}{\Gamma\left(  1-s-n\right)  }\tau^{-s-n}=\frac{1}{\Gamma\left(
1-s\right)  }\frac{\partial^{n}}{\partial\tau^{n}}\tau^{-s}%
\end{equation}
one may rewrite series for $\tau\rightarrow+\infty$%
\begin{equation}
\left(  x+i\varepsilon\right)  ^{s-1}\exp\left\{  -ix\tau\right\}  =2\pi
\sum_{n=0}^{\infty}\delta^{\left(  n\right)  }\left(  x\right)  \frac
{\exp\left\{  i\frac{\pi}{2}\left(  s-1-n\right)  \right\}  }{n!\Gamma\left(
1-s-n\right)  }\tau^{-s-n}%
\end{equation}
as (see \cite{brych})%
\begin{equation}
\left(  x+i\varepsilon\right)  ^{s-1}\exp\left\{  -ix\tau\right\}  =2\pi
\sum_{n=0}^{\infty}\delta^{\left(  n\right)  }\left(  x\right)  \frac
{\exp\left\{  i\frac{\pi}{2}\left(  s-1-n\right)  \right\}  }{n!\Gamma\left(
1-s\right)  }\frac{\partial^{n}}{\partial\tau^{n}}\tau^{-s}%
\end{equation}
and from $\left(  \ref{Mell-Four}\right)  $ we obtain
\begin{equation}
f_{+}\left(  x\right)  \exp\left\{  -ix\tau\right\}  =2\pi\sum_{n=0}^{\infty
}\delta^{\left(  n\right)  }\left(  x\right)  \frac{i^{-n}}{n!}\frac
{\partial^{n}}{\partial\tau^{n}}\frac{1}{2\pi i}\int_{\sigma-i\infty}%
^{\sigma+i\infty}m_{+}\left(  s\right)  \tau^{-s}ds
\end{equation}
that with $\left(  \ref{Mell-Inverce}\right)  $ leads to $\left(
\ref{ba+}\right)  $. Mutatis mutandis the similar result may be obtained for
$f_{-}\left(  x\right)  \exp\left\{  -ix\tau\right\}  $.

\end{document}